\documentclass{article}
\usepackage[english]{babel}
\usepackage[margin=1in]{geometry}
\usepackage{multicol}
\usepackage{graphicx}
\usepackage{wrapfig}
\usepackage{csquotes}
\usepackage{yfonts}
\usepackage[T1]{fontenc}
\usepackage[math]{blindtext}
\usepackage{caption}
\usepackage[framemethod=tikz]{mdframed}

\mdfsetup{%
  middlelinecolor=blue,
  middlelinewidth=1pt,
  roundcorner=10pt,
  leftmargin=1.5cm,
  rightmargin=1.5cm,
  skipabove=1mm,
  skipbelow=1mm
}
  
\begin{document}

{\centering

  {\bfseries\Large A simple numeric algorithm for ancient coin dies identification\bigskip}
  
  Luca Lista\textsuperscript{1} \\
  {\itshape
    \textsuperscript{1} Senior Researcher at Istituto Nazionale di Fisica Nucleare (INFN), Sezione di Napoli (Italy)\\
  }
}
{\centering
  Dated: \today
}
\begin{abstract}
  A simple computer-based algorithm has been developed to identify pre-modern coins
  minted from the  same dies, intending mainly coins minted by hand-made dies
  designed to be applicable to images taken from auction websites or catalogs.
  Though the method is not intended to perform a complete automatic classification,
  which would require more complex and intensive algorithms accessible to experts of computer vision
  its simplicity of use and lack of specific requirement about the quality of pictures can provide 
  help and complementary information to the visual inspection, adding quantitative measurements
  of the ``distance'' between pairs of different coins. The distance metric is
  based on a number of pre-defined reference points that mark key features of the coin to identify the set of coins they have been minted from.
  \bigskip


\end{abstract}

\section{Introduction}

\begin{wrapfigure}{l}{0.5\textwidth}
  \centering
  \includegraphics[width=60mm]{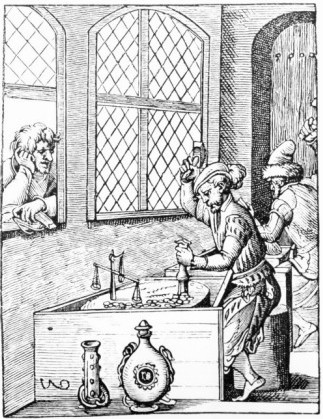}
  \caption{``Der M\"unzmeister'', xylography by Jost Ammanm (1568),
  representing the hand-hammering work in an ancient mint.}
  \label{fig:mint}
\end{wrapfigure}

Before the advent of machine mints, coins were minted by manually hammering
a metallic flan between two dies (Fig.~\ref{fig:mint}).
Dies were replaced with relatively high frequency
due to damages and wear induced by the heavy mechanical stress. For this reason, ancient
and medieval coins of the same type often exhibit visible differences, because they have been 
most likely minted using different dies.
Dies were manually engraved, hence they were different from each other.

The identification
of coins hammered from the same dies (not necessarily the same pair of dies, in
several cases, because not both dies were replaced simultaneously) is of interest for numismatic studies.
Fist attempts to realize die identification using an algorithm date back at least several decades~\cite{Colbert},
and the application of computer-vision techniques to numismatics provides valuable help
to the field (see Ref.~\cite{VAST:VAST08:017-024} and references therein
for a review of the main recent applications).
The full power of computer-based algorithms aims at studying even the most difficult cases,
e.g.: identifying coins in bad conditions or spotting subtle feature of the coins.
But for those purposes, often good quality pictures
(special illumination, high resolution, etc.) or even 3D scans are required.
Sometimes complex algorithms may require large CPU time to analyze coins, which in some
cases may turn out to be a drawback~\cite{Kampel}.

We present here a simple and very fast algorithm based on the manual identification of a small number of reference
points on coin images that turn out to be sufficient to provide quantitative information about
die identification in order to complement the observations performed with a visual inspection.
For the presented application, the choice of reference points may be crucial, and it will depend on the specific identification performed by the analyst of key features of the image.

The method has been applied to a set of medieval coins, namely the golden {\it saluto},
minted under Charles I d'Anjou
(king of Sicily and then king of Naples from 1266 to 1285)~\cite{PR-C1-1} and
Charles II d'Anjou (king of Naples from 1285 to 1309)~\cite{PR-C2-1},
based on a set of images retrieved from public online auctions, catalogs and other websites.

The production of the coin {\it saluto}, later renamed as  {\it carlino}, after the name of 
king Charles I d'Anjou, was established on 18 of April 1278. It was
the first golden coin produced in the mint of Naples. The name {\it saluto} reflects the
scene of the Annunciation of the Virgin, where the angel Gabriel gives his salute
to Mary~\cite{GF}.
The production of this coin continued under Charles II who ordered in 1295 to replicate
the same image commissioned by his father and predecessor, Charles I:
\begin{displayquote}
{\it Volentes indempnitati dicte Curie precavere hoc tibi oretenus et expresse commisso tenore presentium tue
fidelitati mandamus quatenus circa id diligenti consideratione prohabita
diligentia et cautela que honori nostro et comodo videris convenire
intendas et intendi facias per viros idoneos et fideles ac expertos 
intalibus ad cudi faciendam karolenos aureos et argenti qui sint 
illins tenute ac forme sicut erant illi qui cudebantur tempore clare
 memorie domini patris nostri et laborandum in sicla predicta eo modo 
et forma quibus habes ut predicitur oretenus in mandatis circa predicta 
operosus et sollers existens cum consilio magistrorum rationalium.}~\cite{ASNa}
\end{displayquote}

\section{Input images}

Fifty four images of coins from online auction sites and other websites have been considered.
The image resolution ranges
from $280\times 280$ to $1634\times 1634$ pixels, most of the images being in the
range $500\times 500$ to $700\times 700$ pixels.
The list of images and their provenance is reported in the Appendix~\ref{appendix}.

Some of the images have been found to belong to the same coin proposed under multiple
auctions. Those duplicates have not been discarded and have been used, {\it a posteriori},
to study the ability of the method to identify multiple images of the same coin.

\section{Analysis method}

The analysis method is based on the identification of a number of reference points
whose positions on the images are measured and stored to disk. After this first phase, a
numerical analysis is performed to determine coin metrics.

\subsection{Reference points}

\begin{figure}[htbp]
  \centering
  \includegraphics[width=80mm]{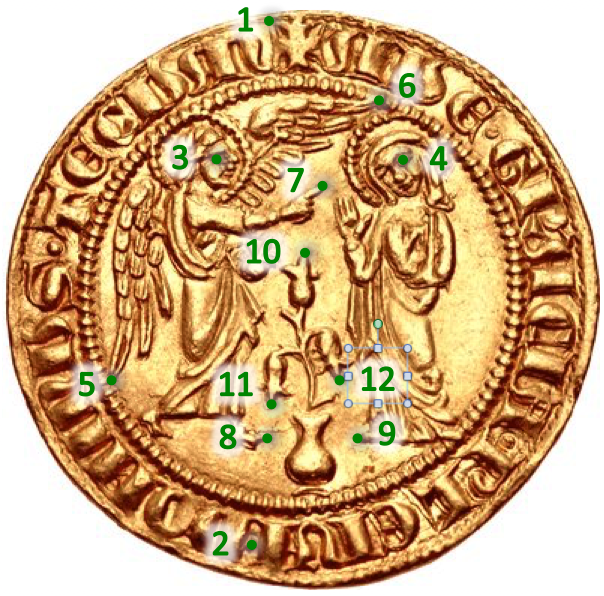}
  \caption{Reference points identified on one coin image. The coin is from CNG, auct. 88, lot 1920.}
  \label{fig:points}
\end{figure}

Twelve reference points are identified on the obverse side of each coin, which
has been chosen because of its distinctive features.
The coin obverse represents the Virgin Mary and the angel Gabriel in the Annunciation scene
(Fig.~\ref{fig:points}).
The points have been sampled manually for all coins
using the open-source application GIMP~\cite{GIMP} and
the corresponding array of points is saved to disk in Scalable Vector Graphics ({\sc svg}) format~\cite{SVG},
which contains, encoded in an {\sc xml}/{\sc ascii} document, the coordinates of each point. The {\sc svg} files have
been later parsed using a C++ code in order to extract the coordinates and analyze them.
The points represent a number of features of the obverse coin side, namely:
\begin{enumerate}
  \item[1] upper corner of left arm of the cross in the top part of the inscription
  \item[2] dot in the bottom part of the inscription between the words ``\textgoth{PLENA}'' and ``\textgoth{DOMINUS}''
  \item[3, 4] right eye of the angel, left eye of the Virgin
  \item[5, 6] extreme points of the angel's wings feathers
  \item[7] angel index finger extreme point
  \item[8, 9] angel and Virgin's feet extremes
  \item[10--12] flowers pistils extremes
\end{enumerate}
In most of the cases, except for the lowest resolution files, the pixel sampling
is finer than the uncertainty due to the manual identification of the points,
which could also be affected, apart from subjective effects, by picture lightening, shadows, etc.

The reference points are defined in each image by a set of coordinate pairs identified as:
\begin{equation}
  \vec{x}_i = (x_i, y_i),\,\, i = 1, \cdots, n\,,
\end{equation}
with $n=12$ being the number of sampled points. 

The choice of reference points is clearly a crucial task of this method, and depends on the choice of the analyst. Points should be easy to determine unambiguously, so vertices of figures should be preferred, and points along curves (e.g.: along a face profile) should be avoided. Moreover, choosing too many points that are very close each other may reduce the discrimination power of the method, due to the finite precision of the image sampling. 

The manual identification of reference points is clearly a somewhat time-consuming task, and could not scale for studies involving large numbers of coins. This may be addressed with automated feature identification, typical of computer-vision algorithms. Anyway, this would limit the usability of the method to expert of computer vision.

\subsection{Relative point distances}

Before attempting any complex algorithm implementation,
one may argue whether a sufficient discrimination of coins minted using different dies
could be achieved inspecting the relative distances of individual pairs of points. In order to take into
account the different magnification of each photo, distances should be 
normalized to a reference unit, which may be conveniently taken as
$D_{12}=|\vec{x_2}-\vec{x_1}|$.
\begin{figure}[htbp]
  \centering
  \includegraphics[width=80mm]{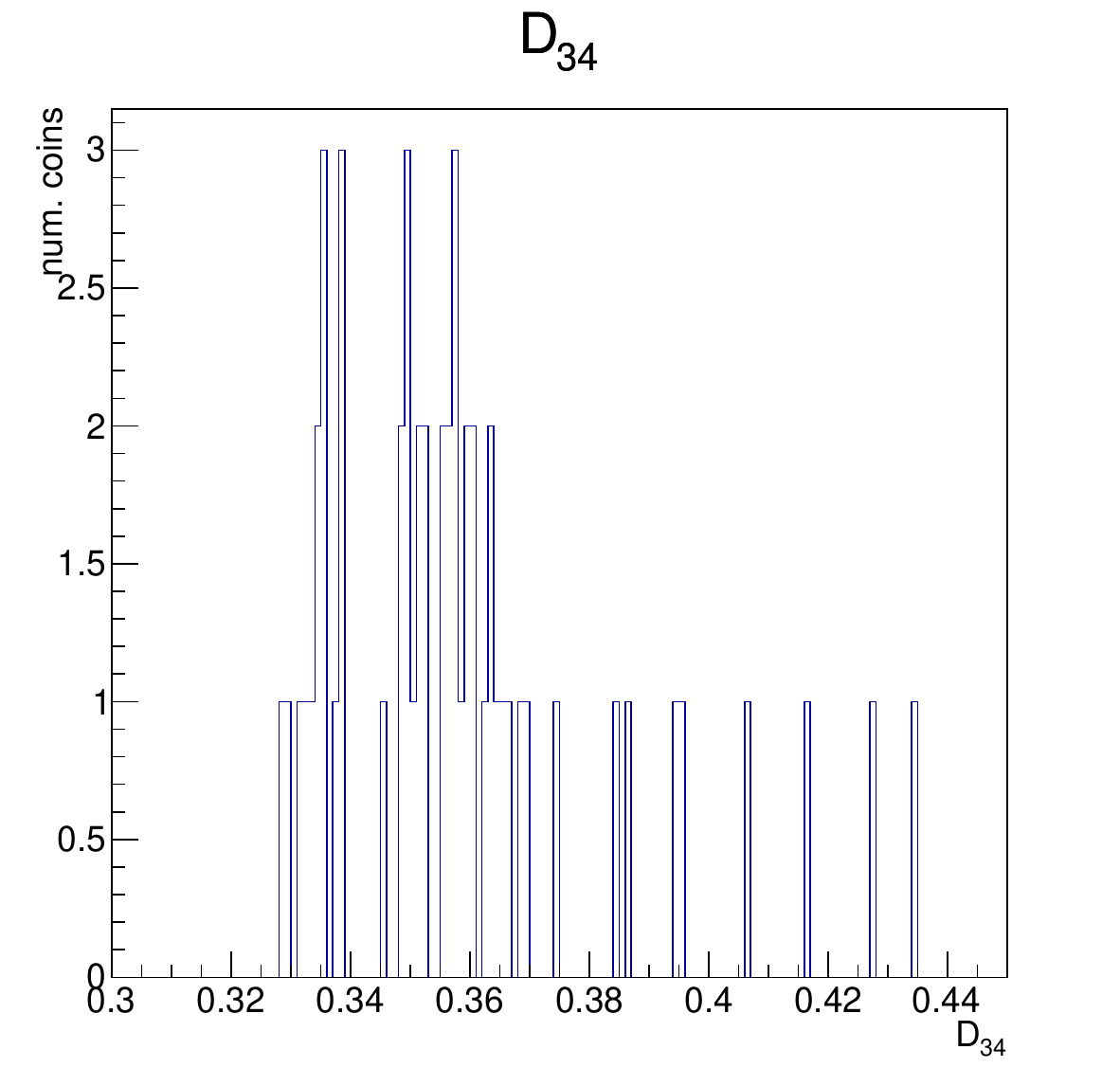}
  \includegraphics[width=80mm]{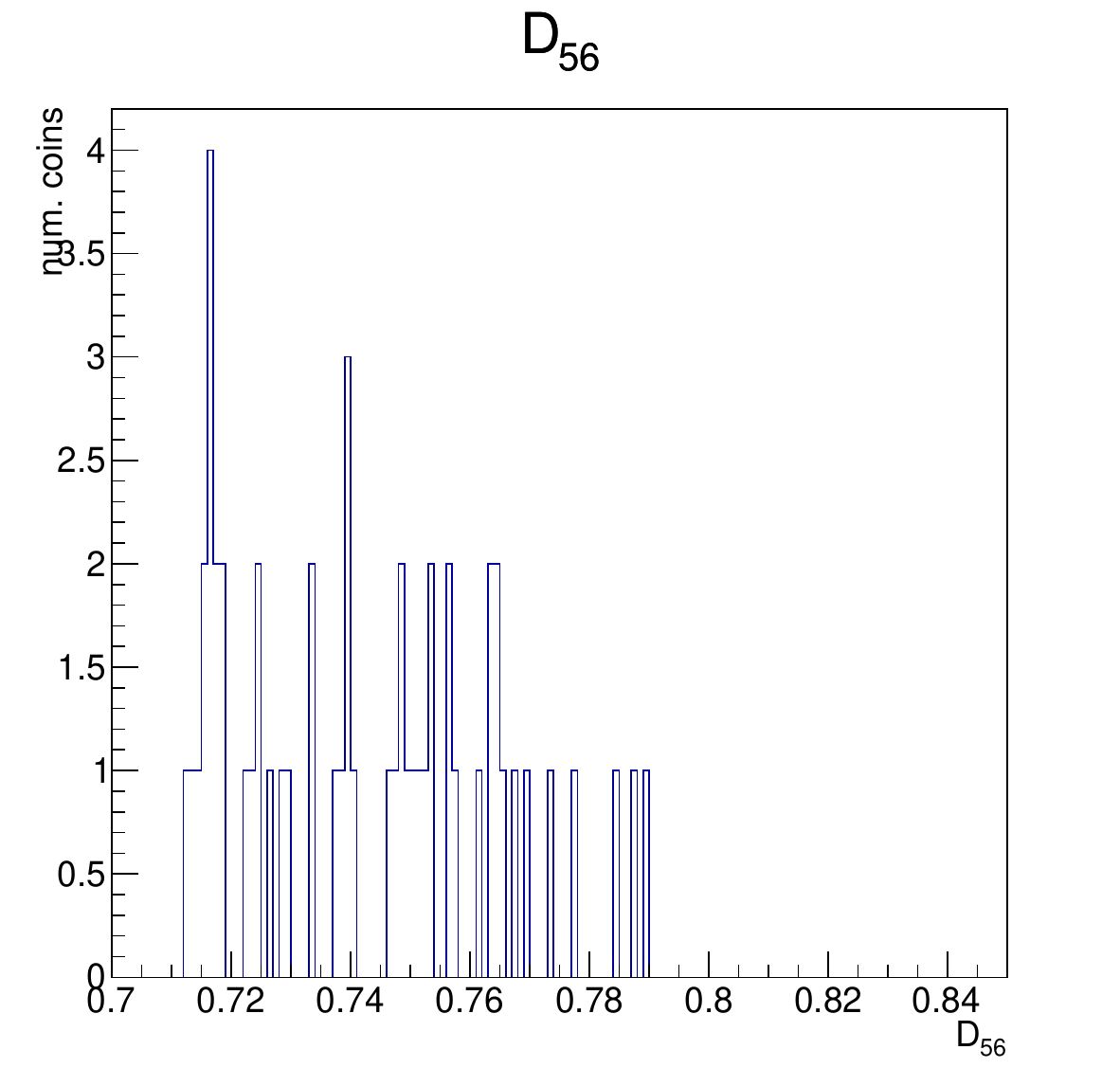}
  \caption{Distribution of $D_{34} =|\vec{x}_3-\vec{x}_4|/|\vec{x}_1-\vec{x}_2|$ and
    $D_{56}=|\vec{x}_5-\vec{x}_6|/|\vec{x}_1-\vec{x}_2|$ for the 54 considered coin images.
     Coins minted from the same die should appear as narrow peaks around the same values in 
    this plot.   }
    \label{fig:cdef}
\end{figure}
If the measured values of:
\begin{equation}
  D_{ij}=|\vec{x}_i-\vec{x}_j|/|\vec{x}_1-\vec{x}_2|\,,
  \label{eq:d_ij}
\end{equation}
given $i$, $j=1, \cdots, n$, $i\ne j$, are determined with sufficient accuracy, coins minted from 
the same die should have values of $D_{ij}$ very close each other, while coins from different
dies would have values of $D_{ij}$ very different from each other.

Figure~\ref{fig:cdef} shows the distribution of $D_{34}$ and
$D_{56}$ for the 54 analyzed coin images. 
A bin width of 0.001 has been chosen for the histograms in order
to possibly identify narrow peaks in the distributions.

Though some clusters of coins accumulate near some particular values
of $D_{34}$ or $D_{56}$, most of the measured values appear very sparse
in the plot, showing that a single measurement doesn't provide a convincing indication 
that allows to unambiguously determine groups of coins belonging to the same die.
\begin{figure}[htbp]
  \centering
  \includegraphics[width=80mm]{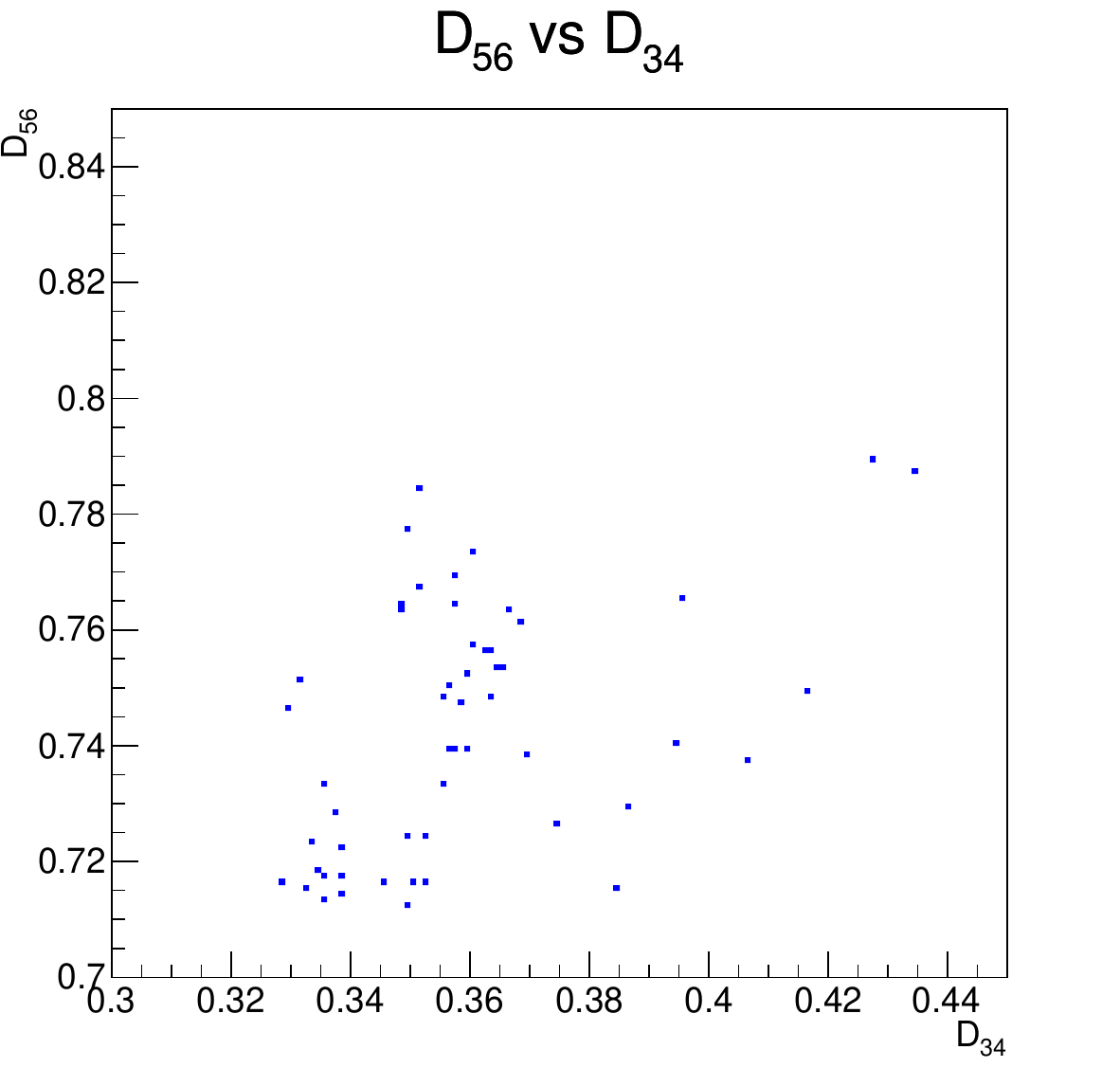}
  \caption{Distribution of $D_{56}=|\vec{x}_5-\vec{x}_6|/|\vec{x}_1-\vec{x}_2|$
    versus $D_{34} =|\vec{x}_3-\vec{x}_4|/|\vec{x}_1-\vec{x}_2|$ for the 54 considered coin images.
    Coins minted from the same die should appear as clusters of points in 
    this plot.
    }
    \label{fig:cdef2d}
\end{figure}
No particular feature can also be spotted in the two-dimensional plot
of $D_{56}$ vs $D_{34}$ (Fig.~\ref{fig:cdef2d}), though the plot
shows some indication of accumulation areas where clusters of coins are present.

The distribution of the relative distances $D_{ij}$ of other coin image pairs, other than
the ones shown in Figs.~\ref{fig:cdef} and~\ref{fig:cdef2d}, and not reported here,
don't show evident features in this sense either.

This motivates the use of a metric which exploits the entire set of
points, as discussed in the following section.

\subsection{Distance metric}

Pictures are taken with different magnification and image resolution, and
a possible tilt of the coin. Those effects are corrected for by using the transformations
described below. Other possible distortions, like perspective effects due to the
camera lens axis not being perfectly aligned with the axis orthogonal to the coin's plane have been
neglected, being second order effect with respect to the misalignment angle.

The applied correction consists of a proper translation, rotation and rescaling
such that for all coins the first two points $\vec{x}_1$ and $\vec{x}_2$ are transformed
into the conventional positions $(0, 0)$ and $(1, 0)$. This is achieved with the following sequence
of transformations of all points $\vec{x}_1,\cdots,\vec{x}_n$:
\begin{enumerate}
\item Translation to the common origin $\vec{x}_1$:
  \begin{equation}
  \vec{x}_i \rightarrow \vec{x}_i - \vec{x}_1\,.
  \end{equation}
  After this translation, the point $\vec{x}_1$ is be transformed into $(0, 0)$.
\item Rescaling of all points using the length $|\vec{x_2}-\vec{x_1}| = |\vec{x_2}|=\sqrt{x_2^2+y_2^2}$
  as unit:
  \begin{equation}
    \vec{x}_i \rightarrow \vec{x}_i / |\vec{x_2}|\,.
  \end{equation}
  After this rescaling, the transformed length
  $|\vec{x_2}-\vec{x_1}|=|\vec{x_2}|$ is equal to unity.
\item Rotation:
  \begin{eqnarray}
    x_i & \rightarrow & s\, x_i + c\, y_i\,,\\
    y_i & \rightarrow & c\, x_i - s\, y_i\,,
  \end{eqnarray}
  where $s = x_2$ and $c = y_2$. After this rotation, the point $\vec{x}_2$ is equal to $(1, 0)$.
\end{enumerate}

In order to compare two coin images, say $A$ and $B$, after the aforementioned set of transformations
has been applied, the following metric is defined to quantify the ``distance'' of the two coin
images:
\begin{equation}
  D^2 = \sum_{i=1}^n |\vec{x}_i^A-\vec{x}_i^B|^2  =\sum_{i=1}^n \left[ (x_i^A-x_i^B)^2 + (y_i^A-y_i^B)^2\right]\,.
\end{equation}
In the sum, the first two terms could be omitted, the difference between first and
second points of $A$ and $B$ being zero, given the application of the above transformation.

The choice of the points 1 and 2 on the coin is motivated by the fact that the two points
are approximately extreme points on the coin and their relative determination suffers
in this way from a small relative uncertainty compared to other nearer points.

Notice that if only four points are identified ($n=4$), then $D^2 = D_{34}^2$, according to Eq.~(\ref{eq:d_ij}).

The idea behind the definition of $D^2$ is that it is equal to a
$\chi^2$ variable, up to a multiplicative factor $\sigma_A^2+\sigma_B^2$:
\begin{equation}
  \chi^2 = \sum_{i=1}^n \frac{(x_i^A-x_i^B)^2 + (y_i^A-y_i^B)^2}{\sigma_A^2+\sigma_B^2} =
  \frac{D^2}{\sigma_A^2+\sigma_B^2}\,,
\end{equation}
where $\sigma_{A,\,B}$ are the uncertainties in the determination of each coordinate of
the coin $A$ or $B$, assumed that all individual coordinates determinations within each coin image have
the same uncertainty.

For the present study a detailed analysis of the uncertainties has not been performed.
A possible extension of the method to properly take into account measurement uncertainties
in order to use a $\chi^2$ variable instead of the variable $D^2$ is discussed in Sec.~\ref{sec:uncert}.

\begin{figure}[htbp]
  \centering
  \includegraphics[width=80mm]{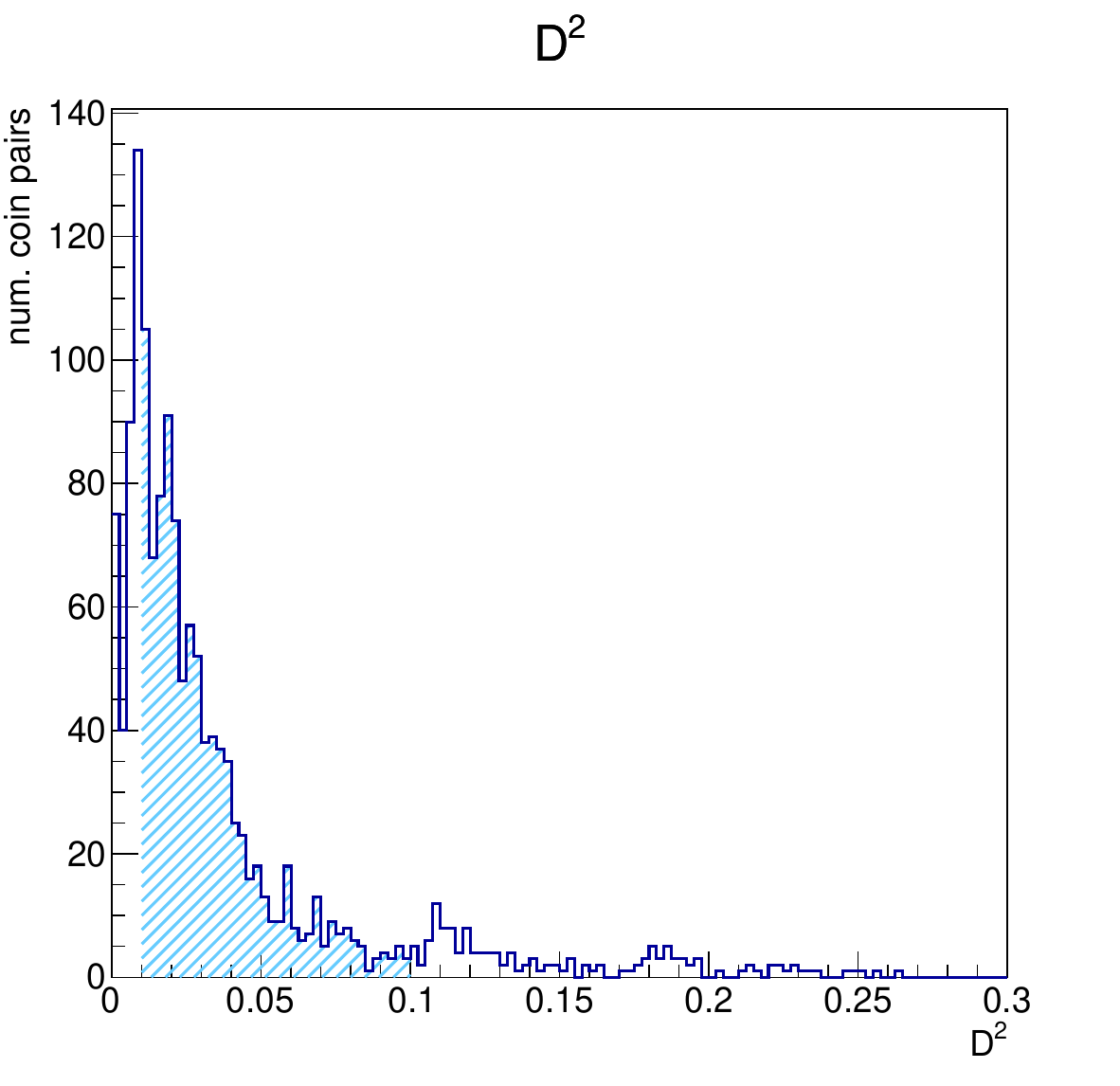}
  \includegraphics[width=80mm]{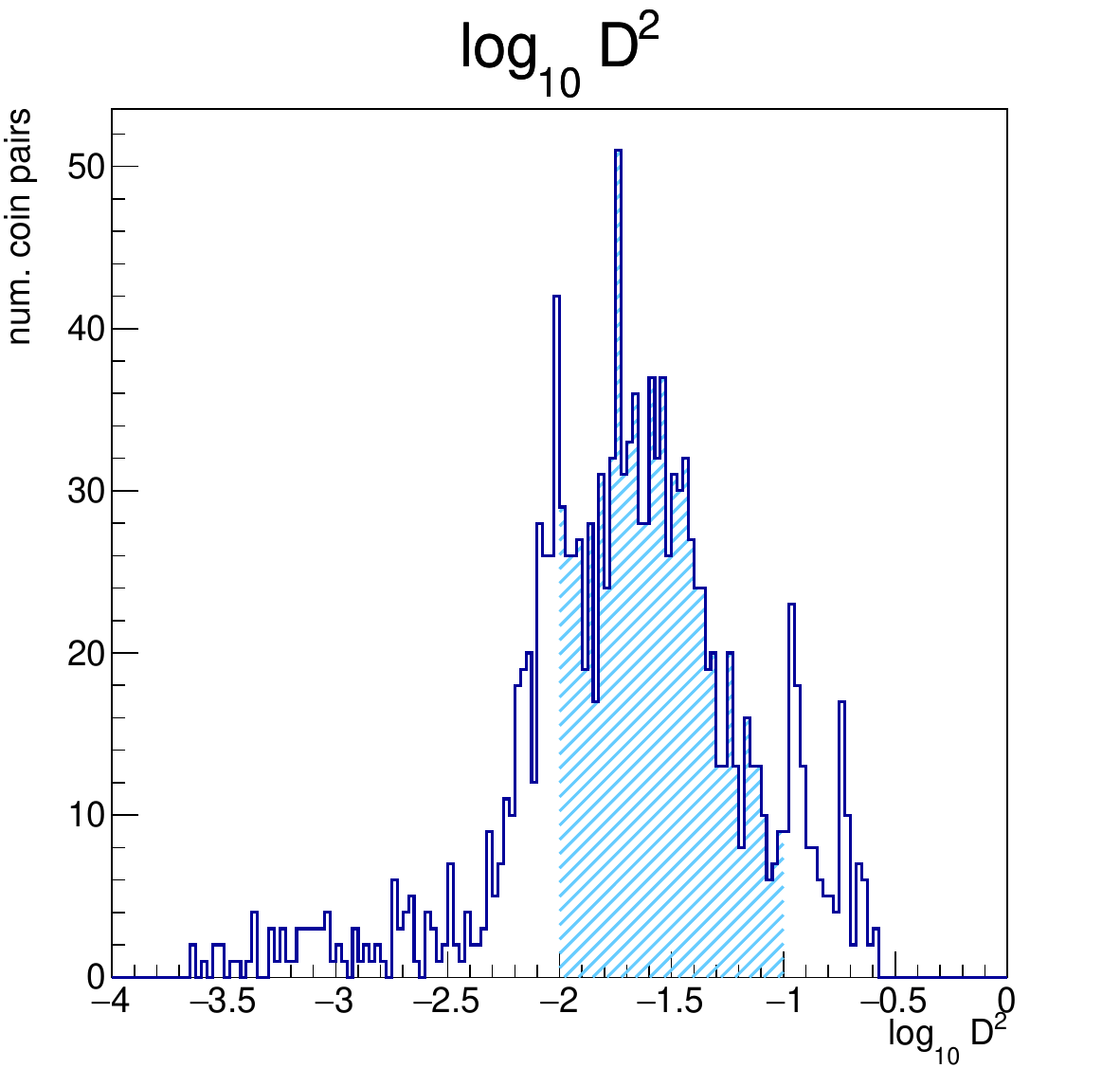}
  \caption{Distribution of $D^2$ in linear (left) and logarithmic (right) scales for the
    54 analyzed coin images. The area corresponding to $0.01<D^2<0.1$ ($-2<\log_{10}D^2<-1$ in
    the right plot)
    is highlighted.
  }
    \label{fig:r2}
\end{figure}

The distribution of $D^2$ for all $54\times53/2=1431$ pairs of the 54 analyzed coin images is shown in
Fig.~\ref{fig:r2} (left). In order to expand the leftmost part of the distribution, where many coin pairs
accumulate, the distribution of $\log_{10}D^2$ is also shown in Fig.~\ref{fig:r2} (right).
Coins minted from the same die have to be identified as pairs having the smallest values of
$D^2$ which appear in the leftmost bins of the histograms
in Fig.~\ref{fig:r2} (left), corresponding to the left side of the histogram
in Fig.~\ref{fig:r2} (right).

\section{Identification of coin dies}

Coin pairs are sorted by increasing value of $D^2$.
Setting a proper threshold on a maximum value of $D^2$ allows to define
pairs of ``compatible'' coins which could most likely belong to the same die. Hence groups of coins all
minted from the same die can be determined accordingly. This procedure was applied together with a visual inspection
of the coin pairs in order to validate the method's results.

\begin{figure}[p]
  \centering
  \begin{mdframed}
  \includegraphics[width=50mm]{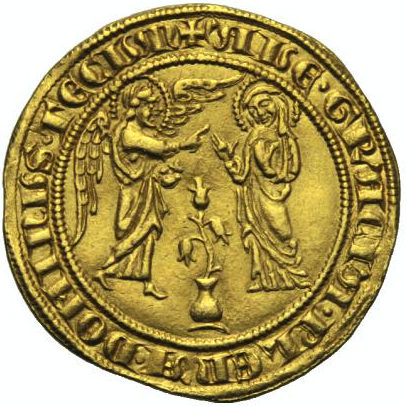}
  \includegraphics[width=50mm]{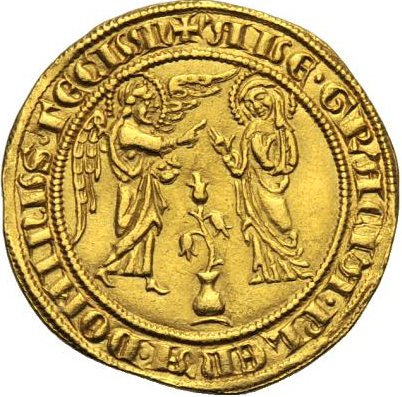}
  $D^2 = 0.000264$
  \end{mdframed}
  \begin{mdframed}
  \includegraphics[width=50mm]{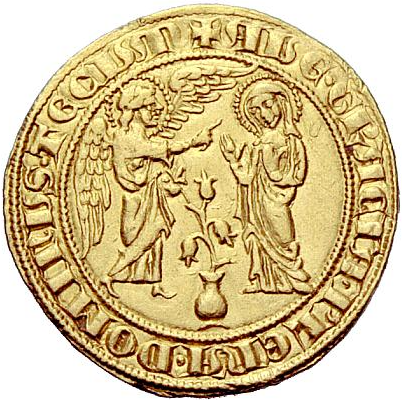}
  \includegraphics[width=50mm]{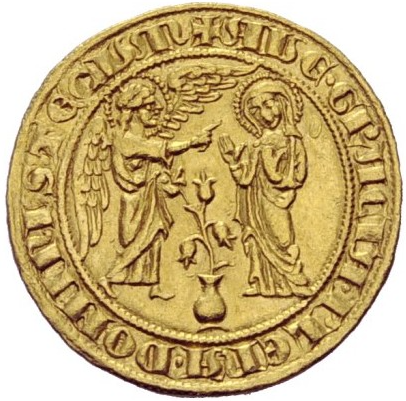}
  $D^2 = 0.000300$
  \end{mdframed}
  \begin{mdframed}
  \includegraphics[width=50mm]{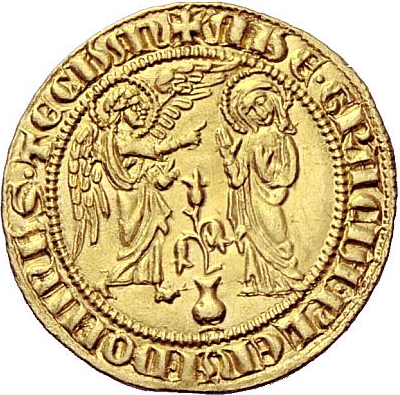}
  \includegraphics[width=50mm]{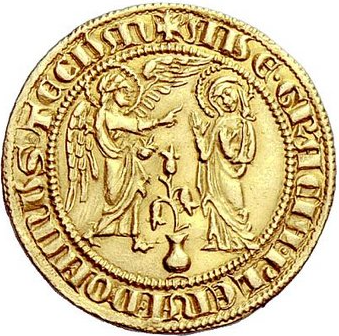}
  $D^2 = 0.00152$
  \end{mdframed}
  \captionsetup{singlelinecheck=off}
  \caption[blah]{Coin pairs with progressively high value of $D^2$.
    \begin{itemize}
    \item $D^2 = 0.000264$: ArsCoin Roma auct. 7, lot 1016 (left); ArsCoin Roma auct. 9, lot 1062 (right); two images of the same coin.
    \item $D^2 = 0.000300$: Hess divo auct. 001, lot 620 (left); Hess divo auct. 326, lot 442 (right).
    \item $D^2 = 0.00152$:  Hess divo auct. 003, lot 764 (left); Hess divo auct. 315, 1203 (right)
    \end{itemize}
  }
  \label{fig:examples}
\end{figure}

\begin{figure}[p]
  \centering
  \begin{mdframed}
  \includegraphics[width=50mm]{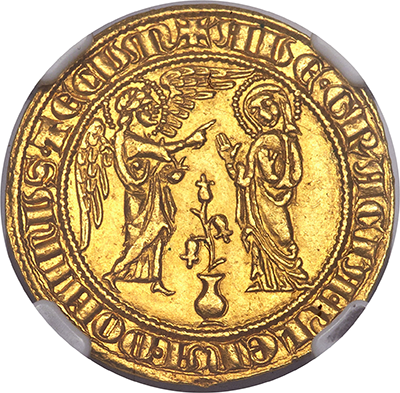}
  \includegraphics[width=50mm]{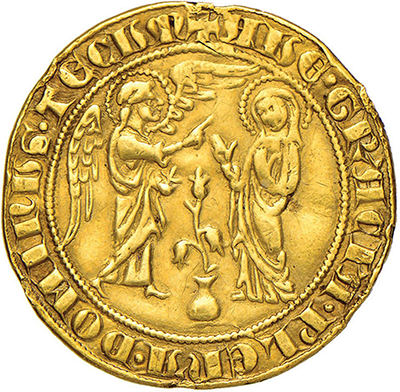}
  $D^2 = 0.00504$
  \end{mdframed}
  \begin{mdframed}
  \includegraphics[width=50mm]{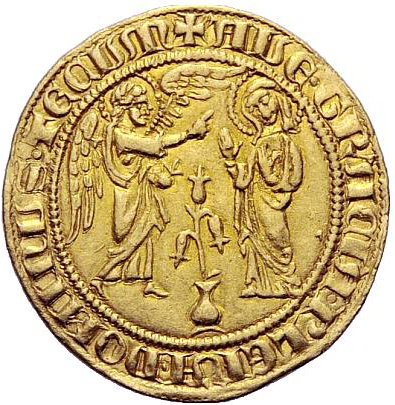}
  \includegraphics[width=50mm]{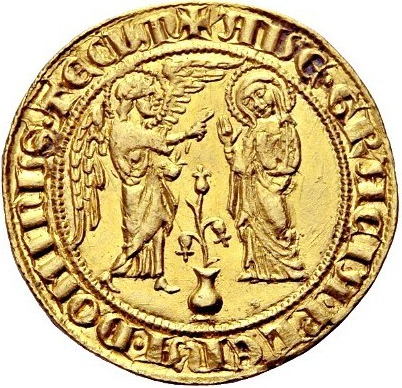}
  $D^2 = 0.00948$
  \end{mdframed}
  \begin{mdframed}
  \includegraphics[width=50mm]{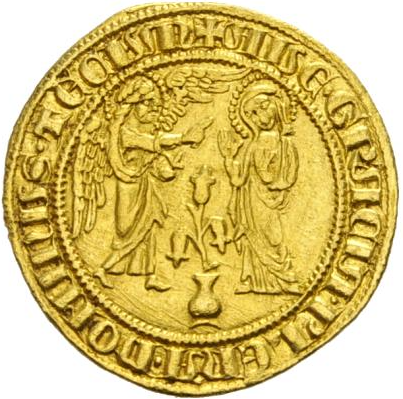}
  \includegraphics[width=50mm]{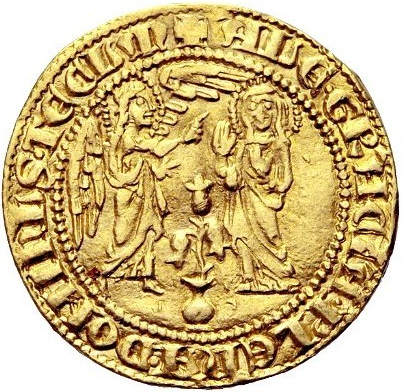}
  $D^2 = 0.0152$
  \end{mdframed}
  \captionsetup{singlelinecheck=off}
  \caption[blah]{Coin pairs with progressively high value of $D^2$.
    \begin{itemize}
    \item $D^2 = 0.00504$: Heritage auct. 3037, lot 31214 (left); Nomisma auct. 52, lot 864 (right).
    \item $D^2 = 0.00971$: Ranieri auct. 3, lot 156 (left); Ranieri auct. 8, lot 284 (right).
    \item $D^2 = 0.0152$: Hess divo auct. 325, lot 468 (left); Ranieri auct. 8, 284 (right). The left coin is from Charles I, the
      right coin is from Charles II.

    \end{itemize}
  }
  \label{fig:examples2}
\end{figure}

Image pairs corresponding to the very lowest values of $D^2$, have
been found to represent the same coin proposed in different auctions. This validates
the power of the method to identify multiple images of the same coin.
Figures~\ref{fig:examples} and~\ref{fig:examples2} show example of image pairs
with increasing values of $D^2$. Images from Fig.~\ref{fig:examples} show coin pairs
which can be visually assigned to the same die (the first pair indeed shows two pictures
of the same coin), while Fig.~\ref{fig:examples2} show coin pairs that exhibit significant
differences at a visual inspection.

All coin pairs having values of $D^2$ lower than $0.0017$ 
appear at a visual inspection to be minted from the same die.
For values of $D^2$ above that threshold, coin pairs apparently belonging to different dies are
found, according to a visual inspection.

\begin{figure}[htbp]
  \centering
  \includegraphics[width=50mm]{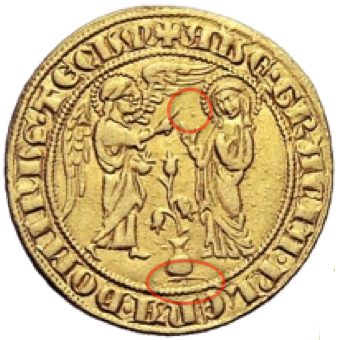}
  \includegraphics[width=50mm]{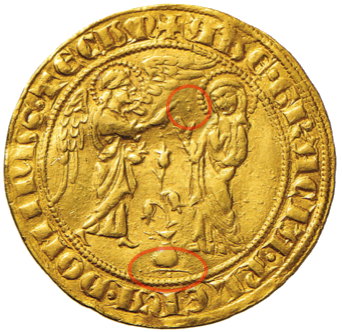}
  \includegraphics[width=50mm]{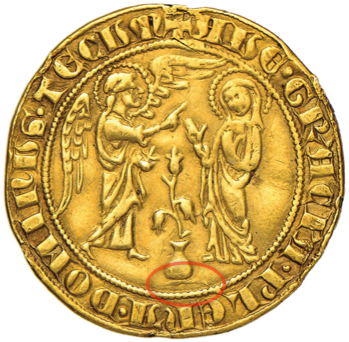}
  \caption{Three coins from Charles I found to be minted from the same die using the metric $D^2$.
    The coins are from Hess divo auct. 305, lot 189 (left, also proposed as
    NAC, auct. 68, lot 240), Inasta auct. 49 lot 1156 (center, also proposed as
    Nomisma, auct. 52, lot 863) and Nomisma auct. 11 online, lot 3331
    (right, also proposed as Nomisma auct. 52, lot 864 and Nomisma auct 53, lot 1142).
    The red circles highlight features, not identified
    as part of the 12 reference points, that appear in multiple coins which confirm that the coins are minted by
    the same die.
  }
    \label{fig:c-i}
\end{figure}

\begin{figure}[htbp]
  \centering
  \includegraphics[width=50mm]{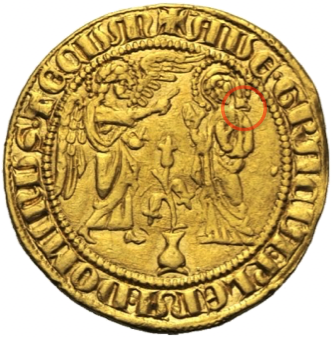}
  \includegraphics[width=50mm]{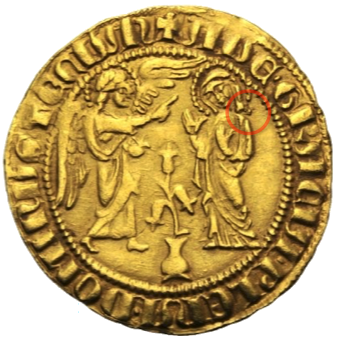}\\
  \includegraphics[width=50mm]{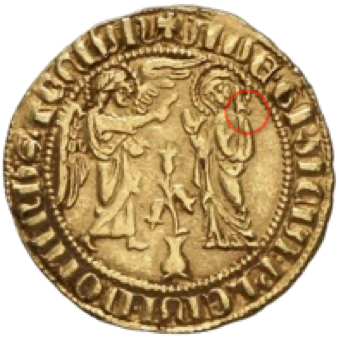}
  \includegraphics[width=50mm]{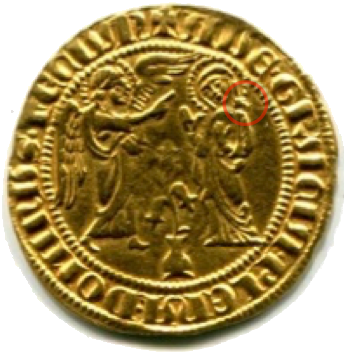}
  \caption{Four coins from Charles I found to be minted from the same die using the metric $D^2$.
    The coins are from ArsCoin Roma auct. 16, lot 935 (top left),
    ArsCoin Roma auct. 12, lot 1092 (top right, also proposed as ArsCoin Roma auct. 16 lot 935),
    Hess divo auction 311, lot 778 (bottom left) and Varesina sale, code 9082 (bottom right).
    The red circles highlight a feature of the Virgin's veil, not identified
    as part of the 12 reference points, that appear all four coins which confirm that the
    coins are minted by  the same die.
}
    \label{fig:c-i2}
\end{figure}

\begin{figure}[htbp]
  \centering
  \includegraphics[width=50mm]{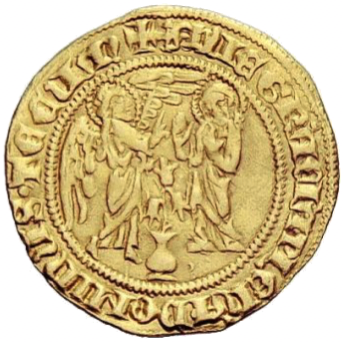}
  \includegraphics[width=50mm]{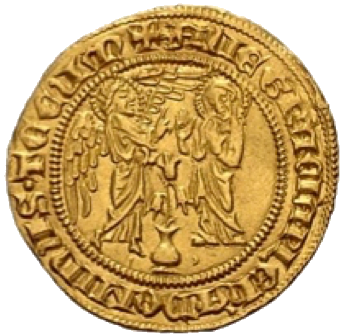}
  \includegraphics[width=50mm]{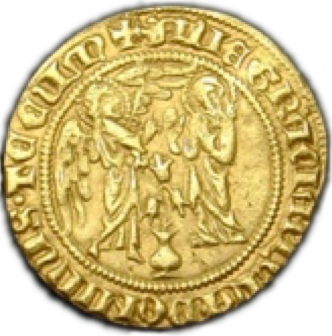}
  \caption{Three coins from Charles II found to be minted from the same die using the metric $D^2$.
    The coins are from NAC auct. 56, lot 937 (left), NAC auct. 35, lot 153 (center)
    and Tkalec, auct. AG, sept. 2008, lot 811 (right).
  }
    \label{fig:c-ii}
\end{figure}

The successful die assignment of the metric $D^2$ is confirmed
by other coin features that could not be identified by the simple sampling
of the 12 reference points, as seen in Fig.~\ref{fig:c-i}, with coins from Charles I.
The coins are from Hess divo auct. 305, lot 189 (also proposed as
NAC, auct. 68, lot 240), Inasta auct. 49 lot 1156 (also proposed as
Nomisma, auct. 52, lot 863) and Nomisma auct. 11 online, lot 3331
(also proposed as Nomisma auct. 52, lot 864 and Nomisma auct 53, lot 1142),
respectively.
Another example is shown in Fig.~\ref{fig:c-i2}, again coins from Charles I.
The coins are from ArsCoin Roma auct. 16, lot 935,
ArsCoin Roma auct. 12, lot 1092 (also proposed as ArsCoin Roma auct. 16 lot 935),
Hess divo auction 311, lot 778 and Varesina sale, code 9082, respectively.
A third example of coins from the same die from Charles II is shown in Fig.~\ref{fig:c-ii}.
The coins are from NAC auct. 56, lot 937, NAC auct. 35, lot 153
and Tkalec, auct. AG, sept. 2008, lot 811,
respectively.

\begin{figure}[htbp]
  \centering
  \includegraphics[width=120mm]{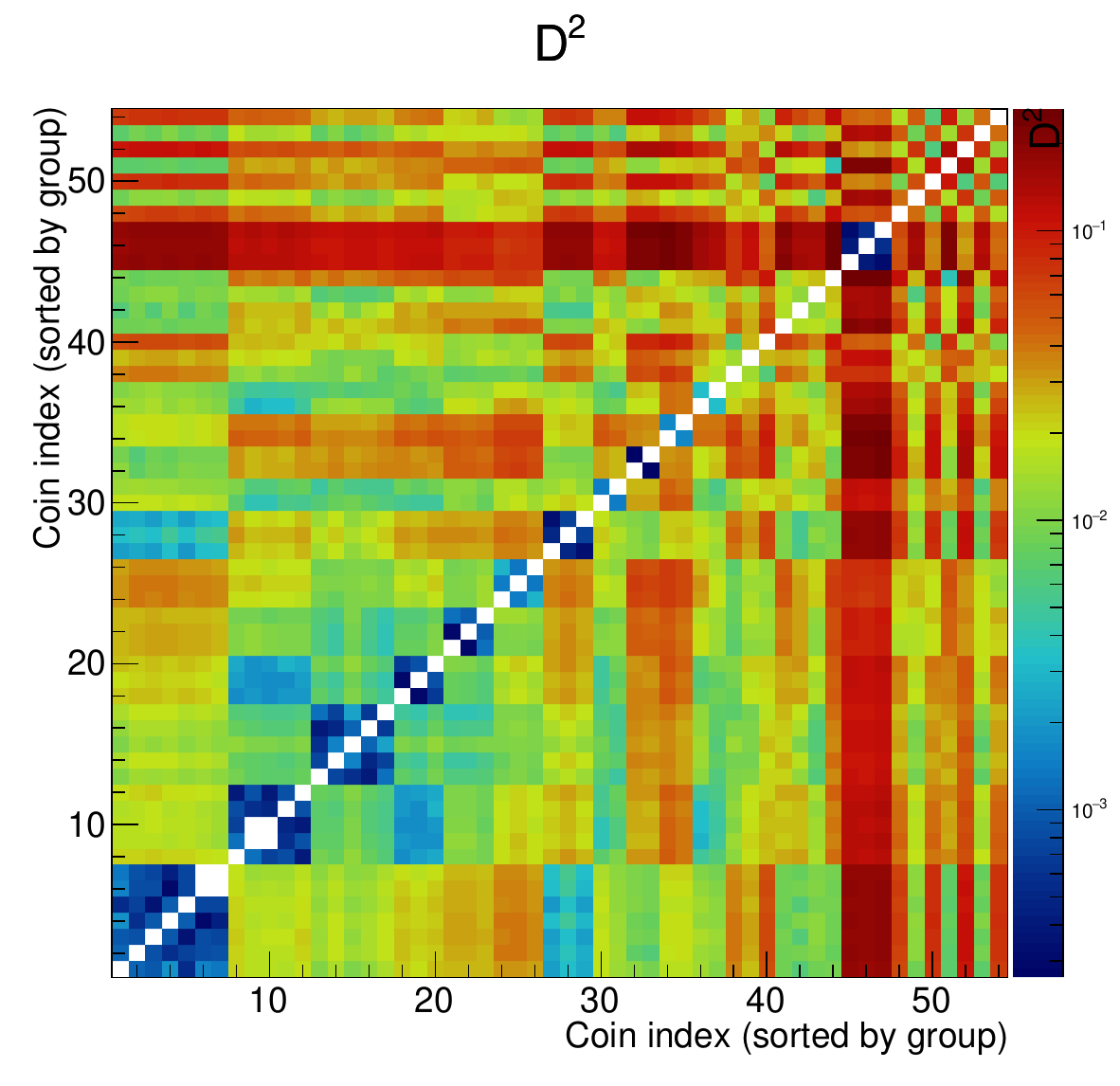}
  \caption{Distribution of $D^2$, represented as color code, for all coin image pairs. Image pairs are
    represented by a dot in two-dimensional plane, where coordinates represent the two images indices.
    Indices are reordered such that images belonging to the same group have adjacent indices.
    Coins from Charles I are put first in the plot (index from 1 to 44), while
    coins from Charles II are put last (index from 45 to 54).
    The bluish square boxes around the white diagonal allow to visually identify the different groups.
  }
  \label{fig:matrix}
\end{figure}
Figure~\ref{fig:matrix} shows the distribution of $D^2$, represented as color code,
for all coin image pairs. Each image pair is represented as a square dot in the two-dimensional
plane whose coordinates are the progressive indices of two images. Images have been reordered by groups
of pairs all having $D^2<0.0017$, such that coins belonging to the same group
have adjacent indices. The plot is symmetric around the diagonal because the pairs
$i-j$ and $j-i$ have the same value of $D^2$, for all indices pairs.
The bluish square boxes around the white diagonal allow to visually identify the different groups.
Coins from Charles I are put first in the scale (index from 1 to 44),
while coins minted under Charles II have been put last in the scale (45 to 54).
A large difference of Charles II coins
with respect to Charles I coins is visible as large bands/areas with red/orange
color, which correspond to $D^2$ around $\sim 100$ times the applied threshold.
This reflects the style changes in the coins from one king to his successor.

The plot also shows similarity in the dies of the second and fourth coin groups,
which anyway can be differentiated by visual inspection by details that are not
sampled by the 12 chosen reference points.
The first pair with a $D^2$ value above threshold, equal to 0.00178,
consists of two coins (Hess divo, auction 311, lot 778 and
Hess divo, auction 325, lot 468) belonging to those two groups.
Those images show very similar features, but also differences are evident at visual inspection,
as visible in Fig.~\ref{fig:c-i-xy}:
\begin{figure}[htbp]
  \centering
  \includegraphics[width=50mm]{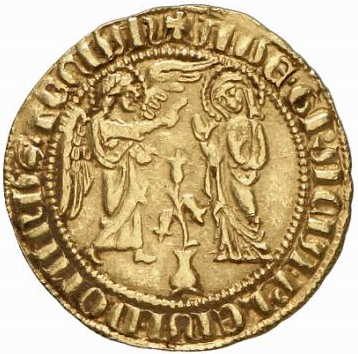}
  \includegraphics[width=50mm]{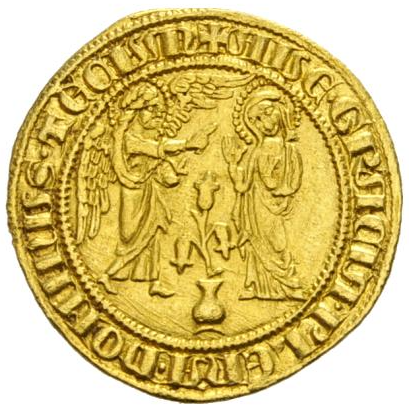}
  \caption{Two coins from Charles I with ``difference'' $D^2=0.001780$ exhibit similarities as well as differences.
    The coins are from Hess divo, auction 311, lot 778 (left) and
    Hess divo, auction 325, lot 468 (right).  }
    \label{fig:c-i-xy}
\end{figure}
the angel's right wing (engraved on the left side of the coin) has the two longest feathers evidently
different in the two coins; the inner feather of the left wing are different, and
the dotted Virgin's nimbus seems to be different as well. This feature could have been spotted
adding extra points on the wings' feathers, but those points have not been added {\it a posteriori}
in order not to bias the method.
Apart from those differences, anyway, the relative position of the inscription compared
to the inner Annunciation image seems very similar in the two pictures. Possibly, the
die has been engraved again after it was worn after frequent usage or otherwise
damaged. But this hypothesis has not
been investigated any further (it would require a more careful analysis), and
the two coin sets have been considered as minted by different dies.
With the exception of the two groups of coins related to those two dies (or possibly
two modified ``versions'' of the same die).
The similarity between the two dies (or possibly the two versions of
the same die) are visible in Fig.~\ref{fig:matrix} as the light blue $3\times 5$ rectangle
between the second group of five coins and fourth group of three coins. 
Apart for those two groups, all other cases show exhibit significant visual differences
and the assignment to different groups seems to been confirmed at a visual inspection.

\section{Results}

Nineteen groups of coins have been identified to be minted by different dies,
for what concerns the coin images from Charles I.
Ten groups contain multiple images, they have 7, 5, 5, 3, 3, 3, 3, 2, 2, 2 images respectively,
a number of which have been identified as the same coin being
proposed in different auctions. The multiple images of the same coin reduce the number of different coins in the
identified groups to 5, 4, 3, 3, 3, 3, 2, 2, 2, 2 respectively. The remaining nine coins
can be considered each belonging to a different die.

For what concerns the coins from Charles II, three coins have been identified as being minted from the same
die, the remaining seven all belong to different dies. No multiple images of the same coin have been found.

All coins have been inspected visually, and the die assignment found by the $D^2$
ordering, provided the assigned threshold of 0.0017, appears to be correct.

\section{Uncertainty analysis and other possible improvements}
\label{sec:uncert}

A proper analysis of uncertainties would improve the definition of the $D^2$
metric, possibly allowing the use of a properly defined $\chi^2$ variable,
whose statistical properties would allow to assign a probability value ($p$-value)
to each coin pair, and, as consequence, a better interpretation of the definition of
a ``distance'' between two coin images. Indeed, the $\chi^2$ definition should take
into account the considered transformation (translation, rotation and scale) in the
error definition and in the computation of the number of degrees of freedom, given
the applied constraint on the position of the first two points, $\vec{x}_1$ and $\vec{x}_2$.

For low-resolution images, an uncertainty contribution due to the pixel sampling
of $1/\sqrt{12}$ times the pixel size should be assigned to the individual measurements
of $x$ and $y$ coordinates of each point. While this contribution
could be negligible for images with high resolution, the present analysis shows
that the metric $D^2$ appears to increase for the lowest-resolution images,
indicating that this contribution should not be neglected in case of poorly
sampled images, which may be a common problem for images taken from websites.

For higher resolution images, a way to estimate the uncertainty
due to the manual identification of the reference points could be done by sampling
multiple images of coins known to be minted from the same die having an
image complexity similar to the one under study. This has not been done
in the present study due to lack of time and lack of image samples,
but could be addressed in the future.

The use of reference points that are determined ``by hand'' can't be accurately
adopted for curves present on the coins' images. Those are the cases with profile faces, which
are frequently displayed on coins. In those cases, reference points can be found
(e.g.: eyes, lips, hairs), but can't exploit the entire information of the image.
Interceptions between such curves and lines joining reference points can be adopted,
but this makes the method less immediate to apply.

The present study assumes two ``privileged'' reference points, numbered as 1 and 2, which
are used to calibrate scale, rotation and translation correction of the images.
In case of incorrect sampling of those two points, uncertainty on the remaining
points is introduced indirectly due to the applied transformations.
In order to avoid this possible effect, the scale, rotation and
translation correction could be performed by applying a best fit to the transformation
parameters minimizing the corresponding $\chi^2$ as a function of those parameters.
This has not been done in the present case, mainly to keep the algorithm simple, and
the implementation easy to reproduce. Such a minimization could anyway be easily implemented using the
{\sc ROOT} framework~\cite{Root}.
The determination of the $p$-value from the
minimized $\chi^2$ should properly take into account the total number of degrees
of freedom of the fit. In case of some missing reference points, due to poor condition of some of the coins,
this could also be treated in the $\chi^2$ definition with a proper accounting of the number of degrees of freedom.
Perspective effects due to misalignment between the camera
lens and axes orthogonal to the coin plane could also in principle be take into account in the fit,
but, as said before, those effects are expected to be small.

The present method is intended to be simple to use, without requiring too much
technological setup or CPU processing time.
Anyway, the power of the method could be improved by the addition of more
than 12 points, at the cost of more manual per-coin intervention.

The use of computer-based image analysis to identify reference points and other
image features would be of great benefit to the application of the method.
Anyway this approach is beyond the scope of the present study, which is intended
to be easily applicable with a small amount of code development and relatively
low effort. On the other hand, the manual detection of reference points avoids possible
failures of automated algorithm in identifying reference points, at the cost of a
longer manual coin inspection time.

\section{Conclusions}

A simple numerical algorithm to identify coins minted using the same die
has been applied on a sample of golden {\it saluto}, a medieval coin minted
under the kings of Naples Charles I and Charles II d'Anjou. The algorithm provides a
measurement of the ``distance'' of two coin images which provides a 
quantitative information that is complementary to the visual inspection of the coins.

The results of the proposed algorithm were consistent with visual inspection:
groups of coins identified as having ``distance'' values below a given threshold
appeared, at visual inspection, to be indeed minted from the same die.

\section{Acknowledgments}
I'm very grateful to Luciano Giannoni for the many e-mail discussions and for
triggering the initiative of the study of Charles I coins. I'm also
grateful to Francesco Di Rauso and Pietro Magliocca for suggesting documentation
and providing me with many hints about the study of
numismatics, in particular about the coinage of Naples kingdom.


\appendix

\section{Appendix}
\label{appendix}
The complete list of images analyzed in this study is reported below. Most
of the images are taken from auction websites, where the action number and lot number
are reported. Few more images have been taken from websites, whose URL is reported.
Images are grouped according to the $D^2$ algorithm described in the text.
This grouping was also used in Fig.~\ref{fig:matrix}.

\vspace{5mm}

{\bf Charles I}

\begin{multicols}{3}
{\footnotesize
  \begin{enumerate}
  \item group:
    \begin{enumerate}
    \item Hess divo 305, 189
    \item \label{inasta_49} Inasta 49, 1156 
    \item NAC 68, 240
    \item \label{nomisma_11} Nomisma 11 online, 3331
    \item Nomisma 53, 1142
    \item[(b$^\prime$)] Nomisma 52, 863 \\(same coin as~\ref{inasta_49})
    \item[(d$^\prime$)] Nomisma 52, 864 \\(same coin as~\ref{nomisma_11})
      \end{enumerate}
  \item group:
    \begin{enumerate}
    \item ArsCoin Roma 12, 1092
    \item \label{arscoin_12} ArsCoin Roma 12, 1093
    \item Hess divo 311, 778
    \item Varesina cod. 9082-d
    \item[(b$^\prime$)] ArsCoin Roma 16, 935\\(same coin as~\ref{arscoin_12})
    \end{enumerate}
  \item group:
    \begin{enumerate}
    \item CNG 88, 1920
    \item Hess divo 003, 764
    \item \label{hessdivo_315} Hess divo 315, 1203
    \item[(c$^\prime$)]  NAC 35, 151 \\(same as coin~\ref{hessdivo_315})
    \item[(c$^{\prime\prime}$)] Rhinocoins,\\ {\scriptsize http://www.rhinocoins.com/ ITALY/rnapsic/CARAN.HTML}\\
      (same as coin~\ref{hessdivo_315})
    \end{enumerate}
  \item group:
    \begin{enumerate}
    \item Hess divo 003, 765
    \item Hess divo 325, 468
    \item Ranieri 4, 316
    \end{enumerate}
  \item group:
    \begin{enumerate}
    \item Artemide XLIII, 610
    \item Heritage 3037, 31214
    \item Hess divo 321, 1316
    \end{enumerate}
  \item group:
    \begin{enumerate}
    \item ArsCoin Roma 5, 901
    \item Artemide XXXIV, 256
    \item CNG 100-1, 788
    \end{enumerate}
  \item group:
    \begin{enumerate}
    \item \label{hessdivo_001} Hess divo 001, 620
    \item Hess divo 326, 442
    \item[(b$^\prime$)] NAC 50, 331\\(same coin as~\ref{hessdivo_001})
    \end{enumerate}
  \item group:
    \begin{enumerate}
    \item ArsCoin Roma 7, 1016
    \item ArsCoin Roma 9, 1052
    \end{enumerate}
  \item group:
    \begin{enumerate}
    \item Ranieri 3, 156
    \item Roma 6, 40
    \end{enumerate}
  \item group:
    \begin{enumerate}
    \item Il portale del Sud,\\ {\scriptsize http://www.ilportaledelsud.org/ monete.htm}
    \item Moneta messinese,\\ {\scriptsize http://monetamessinese.altervista.org/ index.php/storia/13-la-monetazione-di-messina-brevi-cenni}
    \end{enumerate}
  \item Hess divo 323, 489
  \item Nomisma 50, 151\footnote{The resolution of the available image, $350\times 350$, doesn't provide sufficient information for a certain identification.
    By visual inspection, it could be possibly considered compatible with Hess divo 323, 489, but a better image would be needed inorder to ascertain this statement.}
  \item Bolaffi 21, 311
  \item Bolaffi 23, 353
  \item Hess divo 328, 1125
  \item NAC 53, 104
  \item NAC 76, 140
  \item NAC 81, 74
  \item Ranieri 8, 283
  \end{enumerate}
}
\end{multicols}

{\bf Charles II}

\begin{multicols}{3}
  {\footnotesize
    \begin{enumerate}
    \item group:
      \begin{enumerate}
      \item NAC 35, 153
      \item NAC 56, 937
      \item Tkalec AG Sep. 2008, 811
      \end{enumerate}
    \item NAC 57, 206
    \item Bolaffi 21, 312
    \item CNG 72, 2372
    \item Hess divo 001, 621
    \item NAC 76, 139
    \item Ranieri 8, 284
    \item Varesi 67, 265      
    \end{enumerate}
  }
\end{multicols}

\bibliographystyle{ieeetr}
\bibliography{main}

\end{document}